\documentclass[twocolumn,pre, showpacs,preprintnumbers,amsmath,amssymb,letter]{revtex4}
\PassOptionsToPackage{caption=false}{subfig}
\usepackage{graphicx}
\usepackage{dcolumn}
\usepackage{bm}
\usepackage{subfig}

\begin{document}

\title{Effects of Thermal Noise on Pattern Onset in Continuum Simulations of Shaken Granular Layers}

\author{J. Bougie}
 \affiliation{Physics Department,
Loyola University Chicago, Chicago, IL 60626}

\begin{abstract}

The author investigates the onset of patterns in vertically 
oscillated layers of
dissipative particles using numerical solutions of continuum
equations to Navier-Stokes order.  Above a critical accelerational
amplitude of the cell, standing waves form stripe patterns
which oscillate subharmonically with respect to the cell.
Continuum simulations neglecting interparticle friction yield 
pattern wavelengths consistent 
with experiments using frictional particles.
However, the critical acceleration for standing wave formation
is approximately $10\%$ lower in continuum simulations 
without added noise than in molecular
dynamics simulations.  
This report incorporates fluctuating hydrodynamics theory 
into continuum simulations by adding noise terms with 
no fit parameters; this modification 
yields a critical acceleration in agreement
with molecular dynamics simulations.  

\end{abstract}

\date{\today}    

\maketitle

A successful theory of granular hydrodynamics would allow scientists and 
engineers to apply the powerful methods of fluid dynamics to granular flow.  
Despite experimental \cite{bocquet,rericha} and computational 
\cite{ramirez,bougie2005} evidence
demonstrating the potential utility of hydrodynamics models for grains,
a general set of hydrodynamic governing equations is not yet recognized
for granular media \cite{dufty2002,campbell,tsimring}.

One granular hydrodynamics approach derives continuum equations
for number density $n$, velocity ${\bf u}$, and granular temperature 
$T$ ($\frac{3}{2} T$ is the average kinetic energy due
to random particle motion) by modeling particle interactions with 
binary, hard sphere collision operators in kinetic theory 
\cite{goldshtein1,jenkinsandrichman, selaandgoldhirsch}.   
These
equations represent a different approach from other popular methods of
modeling grains, such as molecular dynamics (MD) simulations which
simulate individual grain motion.  
This report is the first to directly incorporate fluctuating 
hydrodynamics theory into continuum simulations of three-dimensional (3D)
time-dependent granular flow. 

Vertically shaken layers provide an important 
testbed for granular phenomena 
\cite{goldshtein2, knight96, brey01, eshuis, melo}. 
A flat layer of grains with depth $H$ oscillated sinusoidally 
in the direction of gravity with frequency $f$ 
and amplitude $A$ leaves the plate
at some time during the cycle if the maximum acceleration 
of the plate $a_{max}=A\left(2\pi f\right)^2$
is greater than the acceleration of gravity $g$.  
Thus the layer
leaves the plate if the dimensionless
accelerational amplitude $\Gamma=a_{max}/g$ exceeds unity.
When $\Gamma$ exceeds a critical value $\Gamma_{C}$, 
the layer spontaneously forms standing waves 
which are subharmonic with respect to the plate.  
Various standing wave patterns are found
experimentally, depending on $\Gamma$ and the dimensionless frequency 
$f^{*}=f\sqrt{H/g}$  \cite{melo}.

Previous experiments \cite{goldman03} and MD simulations \cite{moon03}  
have shown that friction between grains plays a role in these patterns.
Experimentally, adding graphite to reduce
friction decreased $\Gamma_{C}$ and prevented the formation
of stable square or hexagonal patterns found for certain ranges of 
$f^{*}$ and $\Gamma$ in experiments without graphite \cite{goldman03}.   
Similarly, MD simulations with friction between particles have 
quantitatively
reproduced stripe, square, and hexagonal subharmonic standing waves seen
experimentally \cite{bizon98}, but MD simulations without 
friction yield only stable stripe patterns and
display a lower $\Gamma_{C}$ \cite{moon03}.  
In this report, I investigate the onset of stripe patterns in continuum 
simulations of frictionless particles.

Continuum equations for granular media have been proposed using a variety
of approximations 
\cite{bocquet, ramirez, goldshtein1,jenkinsandrichman, 
selaandgoldhirsch, tsimring}.   I use a continuum simulation 
previously used to model shock waves \cite{bougie2002}
and patterns \cite{bougie2005} in a granular shaker in order to directly
compare to previous results \cite{bougie2005}.  
The granular fluid is contained between two
impenetrable horizontal plates at the top and bottom of the container.  
The lower plate oscillates sinusoidally between height $z=0$ and $z=2A$, 
and the ceiling is located at a height $L_{z}$ above the 
lower plate.  
Periodic boundary
conditions are used in the horizontal directions $x$ and $y$ 
to eliminate sidewall effects. 
The dimensions of the box $L_x$, $L_y$, and $L_z$  
can be varied. 
This simulation numerically integrates
equations of Navier-Stokes order proposed by Jenkins and Richman
\cite{jenkinsandrichman} for a dense gas of frictionless
(smooth), inelastic hard spheres with uniform diameter $\sigma$.  
Energy loss due to collisions is characterized by a single parameter, 
the normal coefficient
of restitution $e=0.70$.  
Integrating these
hydrodynamic equations using a second order finite difference scheme 
on a uniform grid in 3D with first order adaptive time stepping 
\cite{bougie2002} yields number density, momentum, and granular
temperature.

Above $\Gamma_{C}$, stripes are seen experimentally for a range of
parameters, including nondimensional frequency $f^{*}=0.4174$, 
and layer depth $H=5.4\sigma$ \cite{melo}.  
In this report, I compare to previous continuum
and MD simulations \cite{bougie2005}, where
$\Gamma$ was varied while frequency $f^{*}=0.4174$ and the number of particles
($6 /\sigma^2$
particles per unit area which experimentally corresponds to a layer depth
$H=5.4\sigma$ as poured \cite{bizon98}) were fixed.  
This corresponds to a frequency of 56 Hz for particles with
diameter $\sigma=0.1mm$.
To compare current results to that previous investigation, 
I use the same frequency, layer depth, 
and cell size
horizontally $L_x = L_y = 42\sigma$ and 
vertically $L_z=80\sigma$ \cite{bougie2005}.

In that report, continuum simulations produced flat layers for 
accelerational amplitudes below
$\Gamma_{C}^{cont} = 1.955 \pm 0.005$, and
stripe patterns above this critical value.
MD simulations produced disordered
peaks and valleys below the onset of stripes, but only displayed 
stripe patterns above  
$\Gamma_{C}^{MD}=2.15 \pm 0.01$, roughly 10\% higher 
than in continuum simulations \cite{bougie2005}.
That study hypothesized that this discrepancy
may be due to fluctuations which were
unaccounted for in the continuum model.

In Rayleigh-B\'{e}nard convection of fluids near the 
onset of convection patterns, fluctuations caused by 
thermal noise create deviations from the dynamics predicted by Navier-Stokes
equations without a noise source.   
Fluctuating hydrodynamics (FHD) theory models these fluctuations 
by adding noise 
terms  to the Navier-Stokes equations 
\cite{landauandlifshitz1959, zaitsev, swifthohenberg}.  FHD
theory accurately describes the dynamics of fluids near
convection onset \cite{wu, oh}.
Experiments indicate that fluctuations due to 
individual grain movement play a larger 
role in granular media than 
do thermal fluctuations in ordinary fluids \cite{goldman2004}.  

Extending FHD theory to granular media
is not trivial.  
The noise terms derived by Landau and Lifshitz \cite{landauandlifshitz1959}
treat fluctuations near equilibrium 
which are small compared to the hydrodynamic fields and
do not provide for local energy loss due to particle inelasticity.
However, granular shaker experiments show
fluctuations much
larger than in ordinary fluids \cite{goldman2004},
and any fluidized granular system is far from equilibrium due to inelastic
particle collisions.  
In the shaken layers considered here, the mean free path
of a particle is on the order of a particle diameter or less, so
fluctuations due to small number statistics may be significant.
Finally, recent simulations of 
a dilute granular gas \cite{brey2009}
showed that Landau-Lifshitz theory underestimates 
fluctuations in a 1D homogeneous cooling state 
by neglecting memory effects of inelastic particles.

As a test of the applicability of FHD, I treat
fluctuations in the granular system analogously to thermal fluctuations
in ordinary fluids.  
At each timestep, the simulation calculates random local stresses
and heat fluxes given by Landau and Lifshitz
\cite{landauandlifshitz1959} at each grid point with no
fit parameters, and includes these terms in the continuum equations 
\cite{jenkinsandrichman, epaps}.

To visualize peaks and valleys formed by standing wave patterns, I calculate
the height of the center of mass of the layer,  $z_{cm}\left( x, y,
t\right)$ as a function of
horizontal
location in the cell at various times $t$.  At a given time $t_0$
and horizontal location $\left( x_0, y_0
\right)$, $z_{cm}\left(x_0, y_0, t_0\right)$ is the
center of mass of all particles whose horizontal coordinates lie within a bin
of size $\Delta x_{bin} \times \Delta y_{bin}$ centered at $\left( x_0, y_0
\right)$.  The simulation grid size defines the bins:
$\Delta x_{bin}=\Delta y_{bin}=2\sigma$.  
Throughout this report, I characterize the patterns at the beginning of a
cycle, when the plate is at its equilibrium
position and moving upwards.
Peaks in the
pattern correspond to maxima of
$z_{cm}$; valleys correspond to minima.

\begin{figure}%
\subfloat{\label{overnonoise}{\includegraphics[width=.205\textwidth]{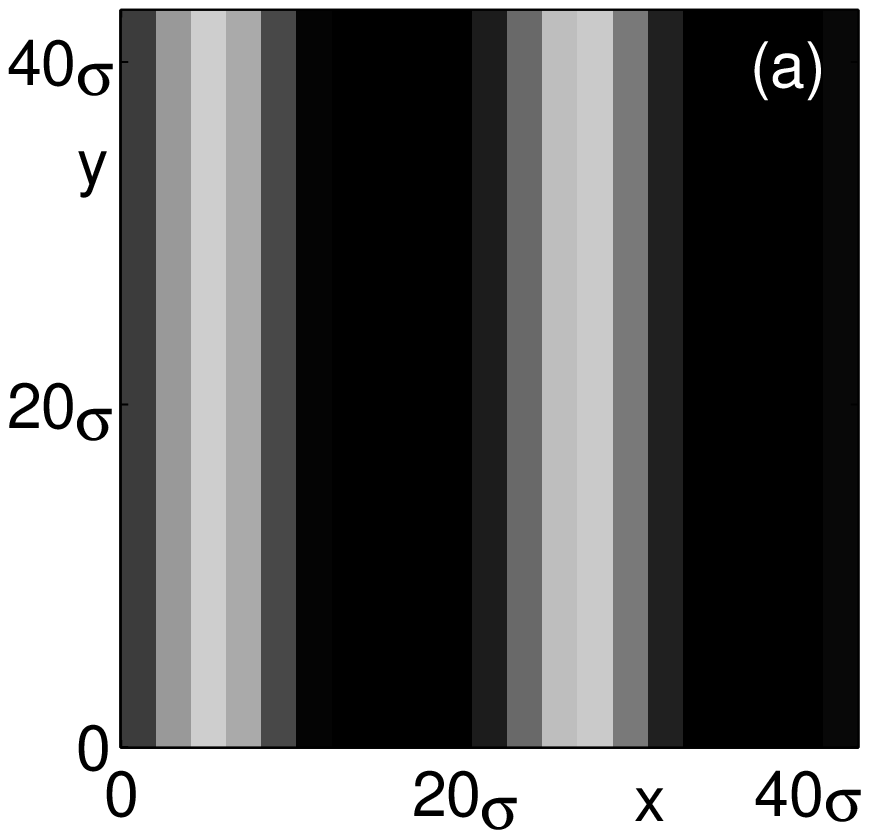}}}
\subfloat{\label{overnoise}{\includegraphics[width=.205\textwidth]{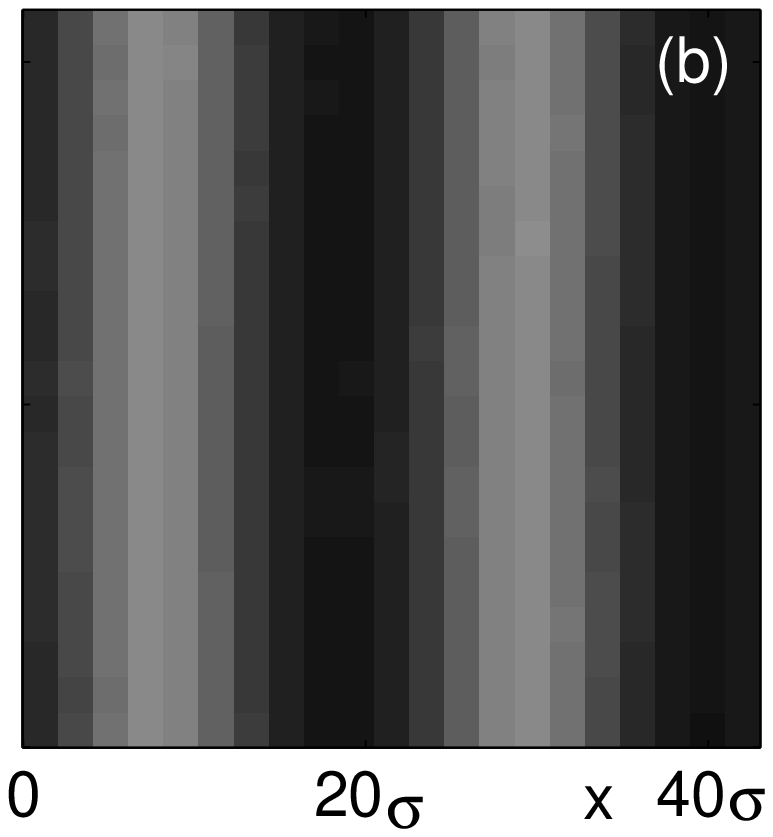}}}\\
\vspace{-0.3cm}%
\subfloat{\label{overcbar}{\includegraphics[width=.35\textwidth]{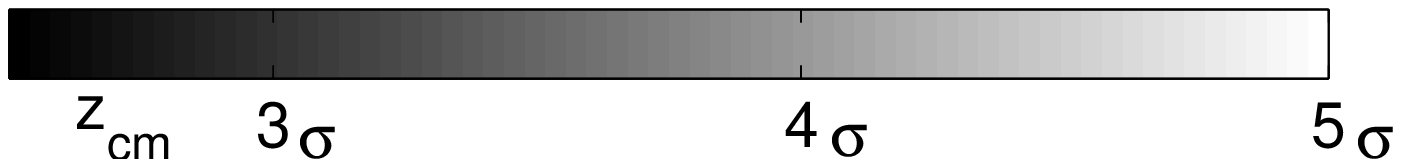}}}
\caption{\label{patterns} An overhead view of a layer of grains, 
  showing the center of mass height $z_{cm}$ as a function of 
  horizontal position $\left( x,y\right)$ in a cell with 
  horizontal dimensions $L_x \times L_y = 42\sigma
  \times 42\sigma$, from continuum simulations (a) without noise 
  and (b) with noise.  
  Peaks of the layer corresponding to large center of mass
  height $z_{cm}$ are shown in white; valleys corresponding to low
  $z_{cm}$ are shown in black.}
\end{figure}

An example standing wave stripe pattern is shown 
in Fig.~\ref{patterns}.  Continuum simulations both
with (Fig.~\ref{overnoise}) and without noise 
(Fig.~\ref{overnonoise}) produce stripe patterns
for $\Gamma=2.2$ and $f^{*}=0.4174$. 
These patterns
oscillate subharmonically, repeating every $2/f$, so the location of
a peak of the pattern becomes a valley after one cycle of the plate, and
vice versa \cite{melo}.  
When the accelerational amplitude is reduced to $\Gamma=1.9$, 
stripes do not appear.  

In both cases,  
two wavelengths fit in the box for this box size and frequency
(Fig. ~\ref{patterns}), although  
simulations without noise 
show sharper peaks and valleys with larger amplitude
than simulations
with noise.
To compare these amplitudes, I examine
the deviation of the height of the
center of mass of the layer as a function of horizontal location
in the
cell from the center of mass height averaged over the entire cell:
\begin{equation}\psi(x,y,t)=z_{cm}(x,y,t)-\left< z_{cm}(x,y,t)
  \right>,\end{equation}  
where brackets represent an average over all horizontal locations 
in the cell at
a given time $t$.   Thus,
$\left< \psi^2(t) \right>$ represents the mean square
deviation of the height of the layer from the mean height of the layer.
Note that $\left< \psi^2 \right>$ is large for layers with high
amplitude
peaks and valleys, and goes to zero as the layer becomes perfectly flat.

To distinguish between ordered patterns (stripes) and disordered 
fluctuations, I characterize the long range order of the pattern.
I first calculate the power spectrum of the pattern
$S\left(k_x,k_y,t\right)$
as a function of wavenumbers $k_x$ and $k_y$.
Transforming to polar coordinates $k_{r}$ and $k_{\theta}$ in $k$ space
and integrating radially yields the
angular orientation of the power spectrum
$S(k_{\theta})$. 
I bin $k_{\theta}$ into 21 bins
between $k_{\theta}=0$ and $k_{\theta}=\pi$, and characterize the long range
order by the fraction of the total integrated power that
lies in the bin
with the maximum power:
\begin{equation}P_{max}=\frac{S_{max}}{\int_{0}^{\pi}S\left(k_{\theta}\right)dk_{\theta}},\label{eq:order}\end{equation}
where $S_{max}$
is the integrated power within an angle $\pi/21$ of the maximum value of
$S(k_{\theta})$.  
For a 
perfectly disordered state, with equal power in all
directions, $P_{max}$ would approach $\frac{1}{21} \approx 0.05$, while 
$P_{max}=1$ for a state with
all power in a single bin.  
Thus $P_{max}$
provides a measure of order when stripes form.

I examine $\left< \psi^2 \right>$ and $P_{max}$ for simulations with varying 
$\Gamma$.  In each case, the simulation begins with a flat layer above the
plate with small amplitude initial random fluctuations.  The simulation runs
for 400 cycles of the plate to reach a periodic
steady state.  Then $\left< \psi^2 \right>$ and $P_{max}$ are averaged over
the next 50 cycles.
Compared to simulations without noise, simulations with noise
show greater variation between cycles in their
final state; I run these simulations
three times for each $\Gamma$ 
to find an average less influenced by transient
behavior.  As patterns occur for $\Gamma=2.20$, but not for 
$\Gamma=1.90$, three additional simulations (for a total of six) 
were run for each $\Gamma$
in the range $1.90\leq\Gamma\leq2.20$ to more precisely locate pattern onset.

\begin{figure}%
\subfloat{\label{nonoisepsisqrd}{\includegraphics[width=.236\textwidth]{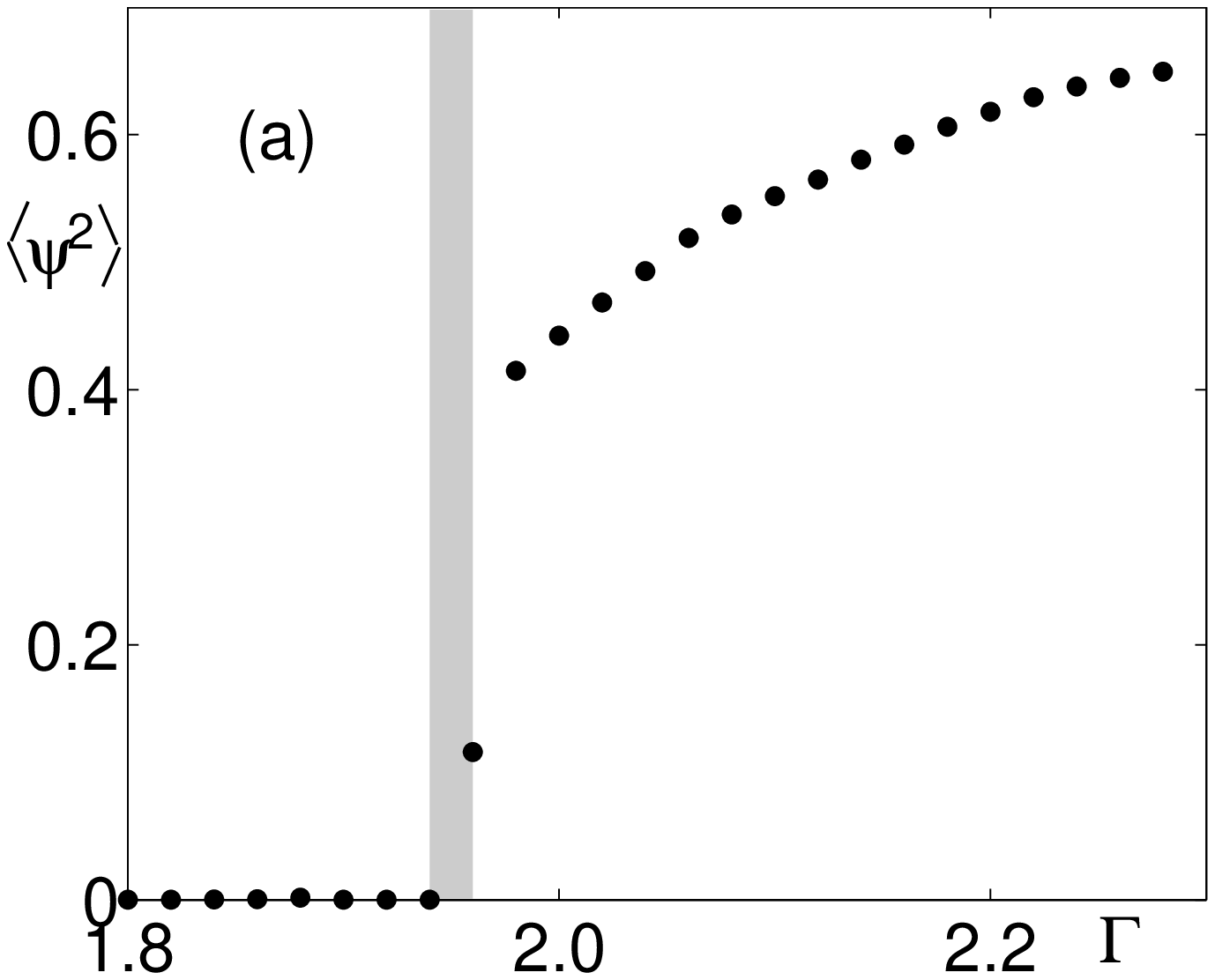}}}
\subfloat{\label{noisepsisqrd}{\includegraphics[width=.227\textwidth]{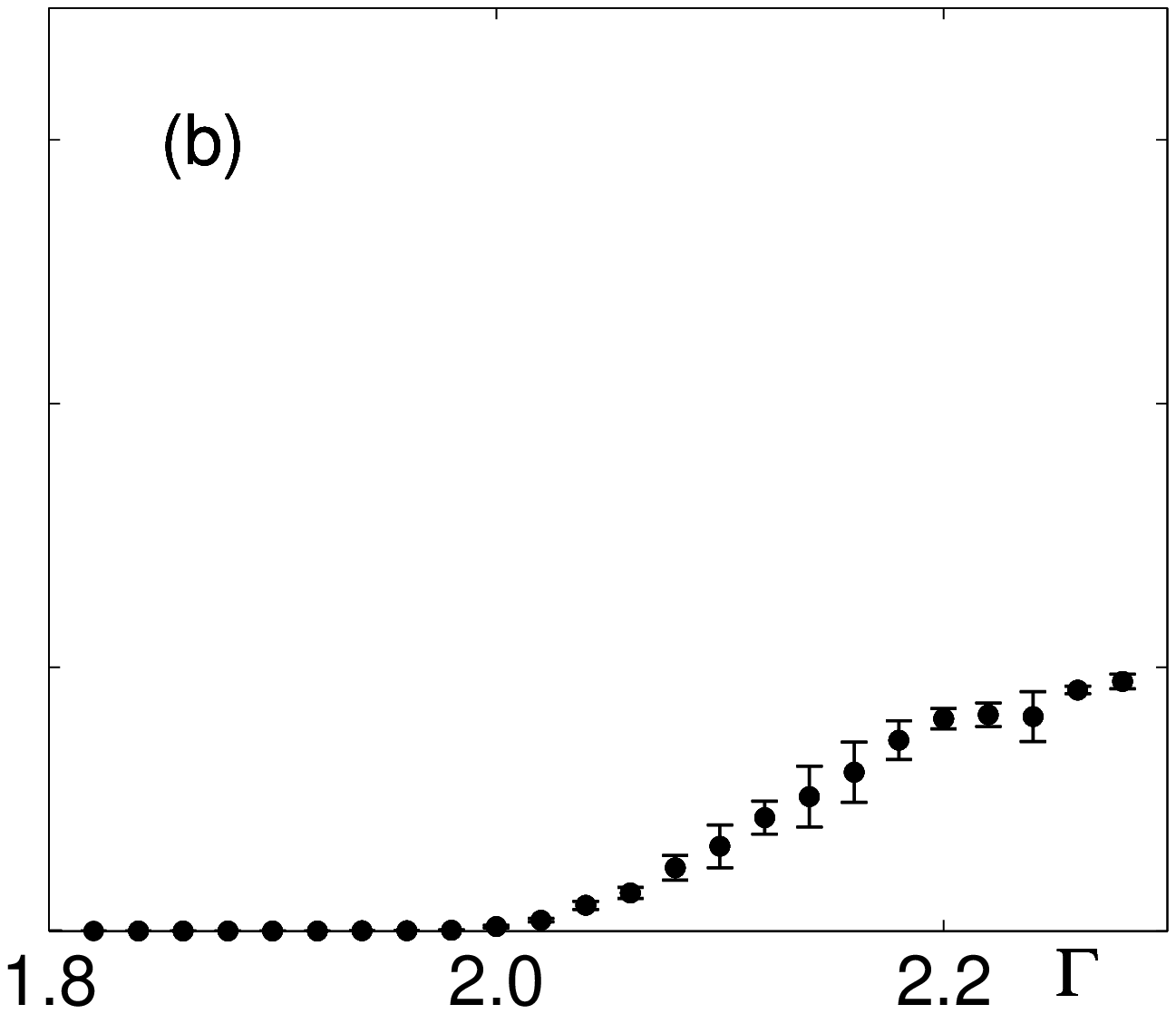}}}
\caption{\label{psisqrd} The mean square deviation 
$\left< \psi^2 \right>$ of the local center of
mass height from the average center of mass height of the layer as a
function of accelerational amplitude $\Gamma$ for simulations
without noise (a), and with noise (b).  In (a),  
$\left< \psi^2 \right>$ is 
averaged over 50 cycles of a single simulation for each $\Gamma$ 
(dots).  
The shaded region $1.94 \leq \Gamma \leq 1.96$ indicates the transition 
between flat layers and layers with non-negligible peaks and valleys.
For simulations with fluctuations, the data in (b)
are averages (dots) with root mean square deviation (bars) 
from
50 cycles from each of six trials 
within the range $1.90 \leq \Gamma \leq 2.20$, and each of three trials
outside that range.}
\end{figure}

For simulations without noise,
fluctuations in the initial condition decay over time for 
$\Gamma \lesssim 1.94$, 
producing a flat layer  (Fig.~\ref{nonoisepsisqrd}).  As $\Gamma$
increases, there is a jump to
a periodic state of non-negligible $\left< \psi^2 \right>$
for $\Gamma = 1.96$, and large amplitude waves occur for all
$\Gamma > 1.96$ (the region $1.94 \leq \Gamma \leq 1.96$ is shaded
in Fig.~\ref{nonoisepsisqrd}).
When noise is added, the layer remains flat for 
some values $\Gamma > 1.96$ (Fig.~\ref{noisepsisqrd}).  
Non-negligible amplitudes of 
$\left< \psi^2 \right>$ occur for $\Gamma \gtrsim 2.0$, but there is not
a sharp jump in amplitude.  

\begin{figure}%
\subfloat{\label{nonoisePmax}{\includegraphics[width=.236\textwidth]{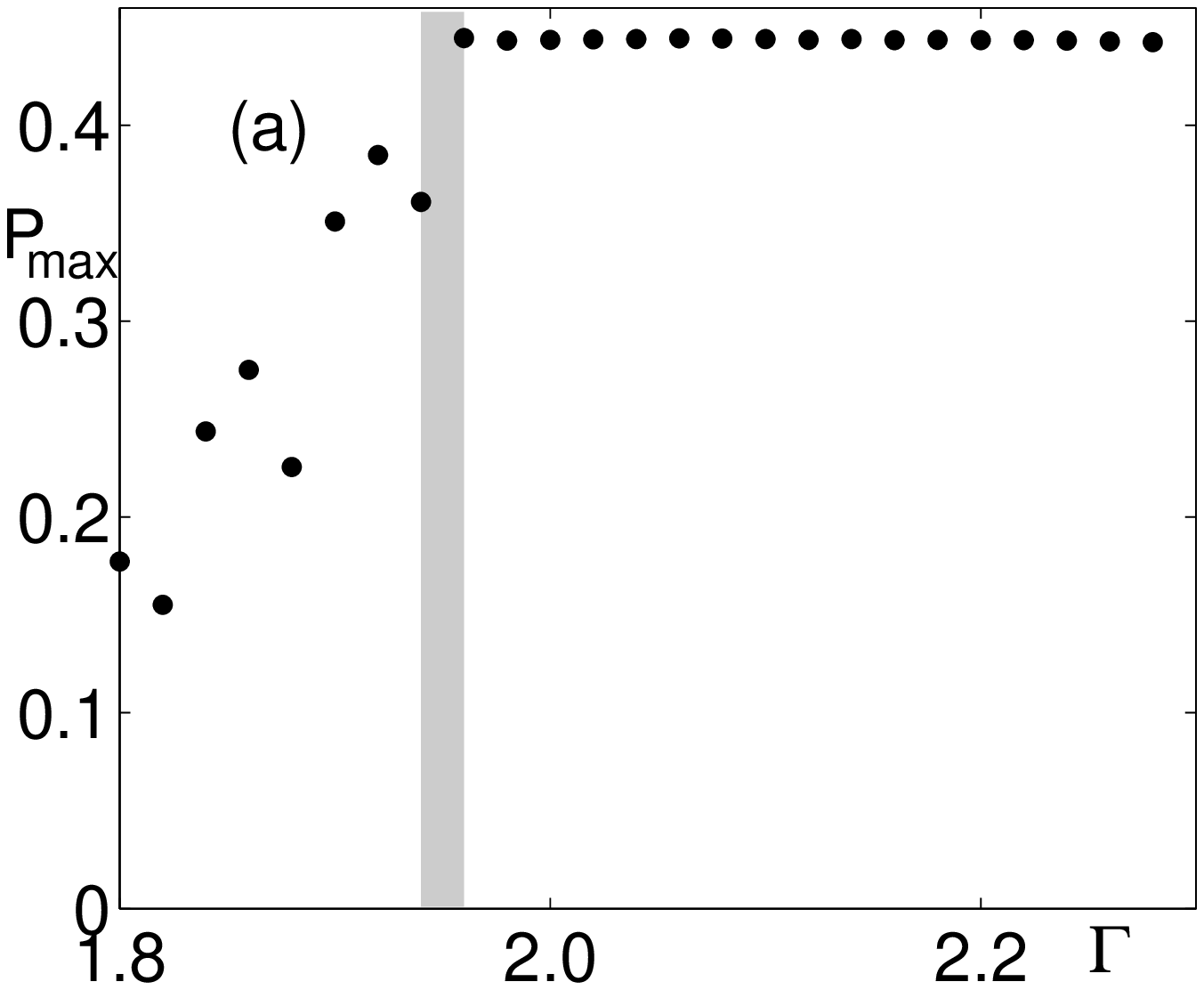}}}
\subfloat{\label{noisePmax}{\includegraphics[width=.227\textwidth]{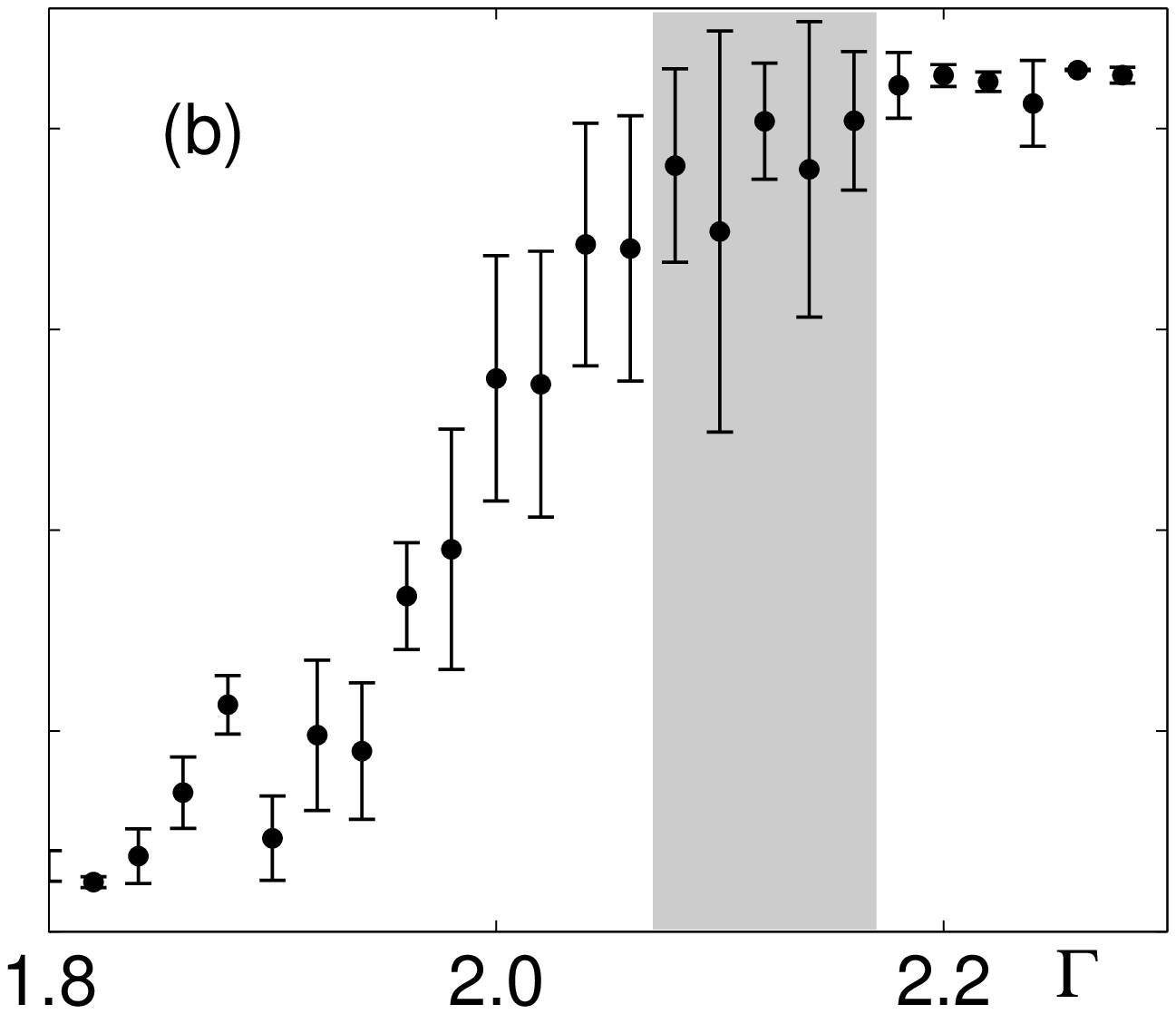}}}
\caption{\label{Pmax} Global ordering $P_{max}$ as a function of 
nondimensional accelerational amplitude $\Gamma$ for continuum simulations
without noise (a), and with noise (b).  For simulations without noise, 
$P_{max}$ is averaged over 50 cycles from a single simulation and represented
as dots, while for simulations with noise, $P_{max}$ is averaged (dots) 
over multiple simulations, with error bars calculated as root mean 
square deviation from this average.  In both cases, there is a
transition (shown in gray) to an approximately constant $P_{max}\approx0.4$.  
The transition area shown in gray is 
$1.94\leq \Gamma \leq 1.96$ in (a) and $2.07\leq\Gamma\leq 2.17$ in (b).}
\end{figure}

Since $\left< \psi^2 \right>$ in Fig.~\ref{noisepsisqrd} 
increases gradually with increasing $\Gamma$ rather than 
showing a sharp onset of waves, I examine the
order parameter $P_{max}$ to distinguish between stripes and 
disordered fluctuations as shown in Fig.~\ref{Pmax}.
For simulations without noise, all layers with
$\Gamma \gtrsim 1.96$ show a nearly constant value of $P_{max} \approx 0.4$
(Fig.~\ref{nonoisePmax}), 
corresponding to the stripe patterns seen in Fig.~\ref{overnonoise}.
For $\Gamma \lesssim 1.94$, the
initial fluctuations decrease over time, leading to a very flat layer 
({\it cf \rm} Fig.~\ref{nonoisepsisqrd}) with lower $P_{max}$. 
I identify the critical value 
$\Gamma_{C}^{cont} =1.95 \pm 0.01$
above which stripe patterns are formed
in simulations without noise. 

For noisy simulations, there is relatively large uncertainty
in $P_{max}$ in the shaded region 
$2.07\lesssim \Gamma \lesssim 2.17$ (Fig.~\ref{noisePmax}).  
Visual inspection shows
transient behavior in this region, with temporary order appearing
and then disappearing, yielding variation in $P_{max}$ 
from simulation to simulation.
Above this shaded region, $P_{max} \approx 0.4$ with
low variation, indicating consistently reproducible stripes.  Below this 
region, $P_{max}$ is consistently lower, indicating disordered fluctuations.
I thus identify the critical value above which stripe patterns form
in simulations with FHD terms
$\Gamma_{C}^{FHD} =2.12 \pm 0.05$.

These results for continuum simulations without noise 
$\Gamma_{C}^{cont} = 1.95 \pm 0.01$ agree with
previous continuum simulations
showing an abrupt transition from a flat layer
to stripe patterns at $\Gamma_{C}^{cont}=1.955 \pm 0.005$ 
\cite{bougie2005}.  
Simulations with
FHD noise, however,
show a gradual increase
of disordered fluctuations below the onset of ordered stripes, and
a transition to stripes at $\Gamma_{C}^{FHD}=2.12\pm0.05$.
While continuum simulations with noise differ from those without
noise, they are consistent with previous MD simulations 
showing the transition to stripe 
patterns at $\Gamma_{C}^{MD}=2.15 \pm 0.01$, with a gradual 
increase in amplitude of disordered fluctuations below this value 
\cite{bougie2005}.

Finally, I investigate the wavelengths of these patterns.
Experiments have shown that wavelength $\lambda$ 
depends on the frequency of oscillation
\cite{melo1993, umbanhowar2000}.  For a range of layer depths
and oscillation frequencies, experimental data for
frictional particles near pattern onset were fit by the
function $\lambda^{*}=1.0+1.1f^{*-1.32\pm0.03}$, where
$\lambda^{*}=\lambda/H$ \cite{umbanhowar2000}.

I investigate frequency dependence by  
holding dimensionless accelerational amplitude $\Gamma=2.2$ constant, 
while varying dimensionless frequency $f^*$.  Simulations
were conducted in a box of size $L_x=168\sigma$,
$L_y=10\sigma$, and $L_z=160\sigma$.  This orientation causes stripes 
to form parallel to the $y-$ axis.  The
dominant wavelength was calculated
from the wavenumber $k_x$
in the $x-$ direction which exhibited the maximum power during
50 cycles of the oscillatory state.  Due to the
periodic boundary conditions and finite box size, wavelengths must
fit in the box an integer number of times, yielding
uncertainty in the wavelength that would be selected in an infinite
box.  

\begin{figure}
\begin{center}
\scalebox{.43}{\includegraphics{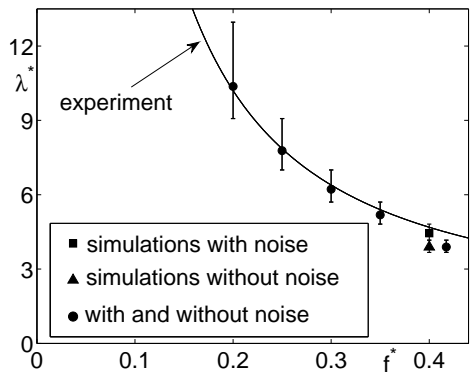}}
\end{center}
\caption{\label{dispersion} 
Dispersion relations for
stripes which form perpendicular to the long dimension of 
cells with horizontal dimensions $168\sigma\times10\sigma$.  Data for
simulations with noise are shown as squares, without noise as triangles,
and points where the two simulations yield the same wavelength are shown as 
circles.
Error bars 
are calculated exclusively from discretization due to
periodic boundary conditions in a finite size box.
In both simulations, the dominant
wavelength of the final oscillatory state $\lambda$ fits very well to the
dispersion relation found in experiments 
$\lambda^{*}=1.0+1.1f^{-1.32\pm0.03}$ (solid line) \cite{umbanhowar2000}.
}\end{figure}

For this box size, frictionless MD simulations and continuum 
simulations without noise have been shown to agree with experimental 
results for frictional particles through the range  
$0.20\lesssim ft \lesssim 0.45$; friction appears
unimportant in wavelength selection through 
this range \cite{bougie2005}.  
Wavelengths found in continuum simulations with and without 
noise are compared to the
dispersion relation fit to experimental data in Fig.~\ref{dispersion}.
Both simulations 
agree quite well with the experimental fit throughout this
range.
The addition of noisy 
fluctuations does not appear to significantly
affect the wavelength of the patterns.

In conclusion, continuum simulations without 
friction can describe important aspects of pattern formation in granular
media.  With or without noise, frictionless continuum simulations 
produce patterns with wavelengths consistent with experimental results
in layers of particles with friction.  
For the shaken layers studied in this report,
patterns in continuum simulations without noise occur for critical 
accelerational amplitude $\Gamma_{C}$ approximately 10\% lower than in 
experimentally verified molecular dynamics simulations.  
Including fluctuating hydrodynamics (FHD) alters the 
onset of patterns; $\Gamma_{C}$ for continuum simulations with noise 
is consistent with MD simulations, but not with continuum simulations
lacking this noise.  
These results indicate that fluctuations play a significant role
in this system, and also suggest directions for further research.
Simulations including memory effects \cite{brey2009} or other 
variations in the FHD model could be compared to test
which approximations significantly alter pattern formation.  
In addition, testing the 
effects of noise on other granular systems will be important in
establishing a general theory of granular hydrodynamics and the role of
fluctuations within that theory.


\begin{thebibliography}{30}
\expandafter\ifx\csname natexlab\endcsname\relax\def\natexlab#1{#1}\fi
\expandafter\ifx\csname bibnamefont\endcsname\relax
  \def\bibnamefont#1{#1}\fi
\expandafter\ifx\csname bibfnamefont\endcsname\relax
  \def\bibfnamefont#1{#1}\fi
\expandafter\ifx\csname citenamefont\endcsname\relax
  \def\citenamefont#1{#1}\fi
\expandafter\ifx\csname url\endcsname\relax
  \def\url#1{\texttt{#1}}\fi
\expandafter\ifx\csname urlprefix\endcsname\relax\def\urlprefix{URL }\fi
\providecommand{\bibinfo}[2]{#2}
\providecommand{\eprint}[2][]{\url{#2}}

\bibitem[{\citenamefont{Bocquet et~al.}(2001)\citenamefont{Bocquet, Losert,
  Schalk, Lubensky, and Gollub}}]{bocquet}
\bibinfo{author}{\bibfnamefont{L.}~\bibnamefont{Bocquet}},
  \bibinfo{author}{\bibfnamefont{W.}~\bibnamefont{Losert}},
  \bibinfo{author}{\bibfnamefont{D.}~\bibnamefont{Schalk}},
  \bibinfo{author}{\bibfnamefont{T.~C.} \bibnamefont{Lubensky}},
  \bibnamefont{and} \bibinfo{author}{\bibfnamefont{J.~P.}
  \bibnamefont{Gollub}}, \bibinfo{journal}{Phys. Rev. E}
  \textbf{\bibinfo{volume}{65}}, \bibinfo{pages}{011307}
  (\bibinfo{year}{2001}).

\bibitem[{\citenamefont{Rericha et~al.}(2001)\citenamefont{Rericha, Bizon,
  Shattuck, and Swinney}}]{rericha}
\bibinfo{author}{\bibfnamefont{E.~C.} \bibnamefont{Rericha}},
  \bibinfo{author}{\bibfnamefont{C.}~\bibnamefont{Bizon}},
  \bibinfo{author}{\bibfnamefont{M.~D.} \bibnamefont{Shattuck}},
  \bibnamefont{and} \bibinfo{author}{\bibfnamefont{H.~L.}
  \bibnamefont{Swinney}}, \bibinfo{journal}{Phys. Rev. Lett.}
  \textbf{\bibinfo{volume}{88}}, \bibinfo{pages}{014302}
  (\bibinfo{year}{2001}).

\bibitem[{\citenamefont{Ram\'{i}rez et~al.}(2000)\citenamefont{Ram\'{i}rez,
  Risso, Soto, and Cordero}}]{ramirez}
\bibinfo{author}{\bibfnamefont{R.}~\bibnamefont{Ram\'{i}rez}},
  \bibinfo{author}{\bibfnamefont{D.}~\bibnamefont{Risso}},
  \bibinfo{author}{\bibfnamefont{R.}~\bibnamefont{Soto}}, \bibnamefont{and}
  \bibinfo{author}{\bibfnamefont{P.}~\bibnamefont{Cordero}},
  \bibinfo{journal}{Phys. Rev. E} \textbf{\bibinfo{volume}{62}},
  \bibinfo{pages}{2521} (\bibinfo{year}{2000}).

\bibitem[{\citenamefont{Bougie et~al.}(2005)\citenamefont{Bougie, Kreft, Swift,
  and Swinney}}]{bougie2005}
\bibinfo{author}{\bibfnamefont{J.}~\bibnamefont{Bougie}},
  \bibinfo{author}{\bibfnamefont{J.}~\bibnamefont{Kreft}},
  \bibinfo{author}{\bibfnamefont{J.~B.} \bibnamefont{Swift}}, \bibnamefont{and}
  \bibinfo{author}{\bibfnamefont{H.~L.} \bibnamefont{Swinney}},
  \bibinfo{journal}{Phys. Rev. E} \textbf{\bibinfo{volume}{71}},
  \bibinfo{pages}{021301} (\bibinfo{year}{2005}).

\bibitem[{\citenamefont{Dufty}(2002)}]{dufty2002}
\bibinfo{author}{\bibfnamefont{J.~W.} \bibnamefont{Dufty}}, in
  \emph{\bibinfo{booktitle}{Challenges in Granular Physics}}, edited by
  \bibinfo{editor}{\bibfnamefont{T.}~\bibnamefont{Halsey}} \bibnamefont{and}
  \bibinfo{editor}{\bibfnamefont{A.}~\bibnamefont{Mehta}}
  (\bibinfo{publisher}{World Scientific},
  \bibinfo{year}{2002}).

\bibitem[{\citenamefont{Campbell}(1990)}]{campbell}
\bibinfo{author}{\bibfnamefont{C.~S.} \bibnamefont{Campbell}},
  \bibinfo{journal}{Annu. Rev. Fluid Mech.} \textbf{\bibinfo{volume}{22}},
  \bibinfo{pages}{57} (\bibinfo{year}{1990}).

\bibitem[{\citenamefont{Aranson and Tsimring}(2006)}]{tsimring}
\bibinfo{author}{\bibfnamefont{I.}~\bibnamefont{Aranson}} \bibnamefont{and}
  \bibinfo{author}{\bibfnamefont{L.}~\bibnamefont{Tsimring}},
  \bibinfo{journal}{Rev. Mod, Phys.} \textbf{\bibinfo{volume}{78}},
  \bibinfo{pages}{641} (\bibinfo{year}{2006}).

\bibitem[{\citenamefont{Goldshtein and Shapiro}(1995)}]{goldshtein1}
\bibinfo{author}{\bibfnamefont{A.}~\bibnamefont{Goldshtein}} \bibnamefont{and}
  \bibinfo{author}{\bibfnamefont{M.}~\bibnamefont{Shapiro}},
  \bibinfo{journal}{J. Fluid Mech.} \textbf{\bibinfo{volume}{282}},
  \bibinfo{pages}{75} (\bibinfo{year}{1995}).

\bibitem[{\citenamefont{Jenkins and Richman}(1985)}]{jenkinsandrichman}
\bibinfo{author}{\bibfnamefont{J.~T.} \bibnamefont{Jenkins}} \bibnamefont{and}
  \bibinfo{author}{\bibfnamefont{M.~W.} \bibnamefont{Richman}},
  \bibinfo{journal}{Arch. Rat. Mech. Anal.} \textbf{\bibinfo{volume}{87}},
  \bibinfo{pages}{355} (\bibinfo{year}{1985}).

\bibitem[{\citenamefont{Sela and Goldhirsch}(1998)}]{selaandgoldhirsch}
\bibinfo{author}{\bibfnamefont{N.}~\bibnamefont{Sela}} \bibnamefont{and}
  \bibinfo{author}{\bibfnamefont{I.}~\bibnamefont{Goldhirsch}},
  \bibinfo{journal}{J. Fluid Mech.} \textbf{\bibinfo{volume}{361}},
  \bibinfo{pages}{41} (\bibinfo{year}{1998}).

\bibitem[{\citenamefont{Knight et~al.}(1996)\citenamefont{Knight, Ehrichs,
  Kuperman, Flint, Jaeger, and Nagel}}]{knight96}
\bibinfo{author}{\bibfnamefont{J.~B.} \bibnamefont{Knight}},
  \bibinfo{author}{\bibfnamefont{E.~E.} \bibnamefont{Ehrichs}},
  \bibinfo{author}{\bibfnamefont{V.~Y.} \bibnamefont{Kuperman}},
  \bibinfo{author}{\bibfnamefont{J.~K.} \bibnamefont{Flint}},
  \bibinfo{author}{\bibfnamefont{H.~M.} \bibnamefont{Jaeger}},
  \bibnamefont{and} \bibinfo{author}{\bibfnamefont{S.~R.} \bibnamefont{Nagel}},
  \bibinfo{journal}{Phys. Rev. E} \textbf{\bibinfo{volume}{54}},
  \bibinfo{pages}{5726} (\bibinfo{year}{1996}).

\bibitem[{\citenamefont{Brey et~al.}(2001)\citenamefont{Brey, Ruiz-Montero, and
  Moreno}}]{brey01}
\bibinfo{author}{\bibfnamefont{J.~J.} \bibnamefont{Brey}},
  \bibinfo{author}{\bibfnamefont{M.~J.} \bibnamefont{Ruiz-Montero}},
  \bibnamefont{and} \bibinfo{author}{\bibfnamefont{F.}~\bibnamefont{Moreno}},
  \bibinfo{journal}{Phys. Rev. E} \textbf{\bibinfo{volume}{63}},
  \bibinfo{pages}{061305} (\bibinfo{year}{2001}).

\bibitem[{\citenamefont{Eshuis et~al.}(2005)\citenamefont{Eshuis, van~der
  Weele, van~der Meer, and Lohse}}]{eshuis}
\bibinfo{author}{\bibfnamefont{P.}~\bibnamefont{Eshuis}},
  \bibinfo{author}{\bibfnamefont{K.}~\bibnamefont{van~der Weele}},
  \bibinfo{author}{\bibfnamefont{D.}~\bibnamefont{van~der Meer}},
  \bibnamefont{and} \bibinfo{author}{\bibfnamefont{D.}~\bibnamefont{Lohse}},
  \bibinfo{journal}{Phys. Rev. Lett.} \textbf{\bibinfo{volume}{95}},
  \bibinfo{pages}{258001} (\bibinfo{year}{2005}).

\bibitem[{\citenamefont{Melo et~al.}(1994)\citenamefont{Melo, Umbanhowar, and
  Swinney}}]{melo}
\bibinfo{author}{\bibfnamefont{F.}~\bibnamefont{Melo}},
  \bibinfo{author}{\bibfnamefont{P.}~\bibnamefont{Umbanhowar}},
  \bibnamefont{and} \bibinfo{author}{\bibfnamefont{H.~L.}
  \bibnamefont{Swinney}}, \bibinfo{journal}{Phys. Rev. Lett.}
  \textbf{\bibinfo{volume}{72}}, \bibinfo{pages}{172} (\bibinfo{year}{1994}).

\bibitem[{\citenamefont{Goldshtein et~al.}(1995)\citenamefont{Goldshtein,
  Shapiro, Moldavsky, and Fichman}}]{goldshtein2}
\bibinfo{author}{\bibfnamefont{A.}~\bibnamefont{Goldshtein}},
  \bibinfo{author}{\bibfnamefont{M.}~\bibnamefont{Shapiro}},
  \bibinfo{author}{\bibfnamefont{L.}~\bibnamefont{Moldavsky}},
  \bibnamefont{and} \bibinfo{author}{\bibfnamefont{M.}~\bibnamefont{Fichman}},
  \bibinfo{journal}{J. Fluid Mech.} \textbf{\bibinfo{volume}{287}},
  \bibinfo{pages}{349} (\bibinfo{year}{1995}).

\bibitem[{\citenamefont{Goldman et~al.}(2003)\citenamefont{Goldman, Shattuck,
  Moon, Swift, and Swinney}}]{goldman03}
\bibinfo{author}{\bibfnamefont{D.~I.} \bibnamefont{Goldman}},
  \bibinfo{author}{\bibfnamefont{M.~D.} \bibnamefont{Shattuck}},
  \bibinfo{author}{\bibfnamefont{S.~J.} \bibnamefont{Moon}},
  \bibinfo{author}{\bibfnamefont{J.~B.} \bibnamefont{Swift}}, \bibnamefont{and}
  \bibinfo{author}{\bibfnamefont{H.~L.} \bibnamefont{Swinney}},
  \bibinfo{journal}{Phys. Rev. Lett.} \textbf{\bibinfo{volume}{90}},
  \bibinfo{pages}{104302} (\bibinfo{year}{2003}).

\bibitem[{\citenamefont{Moon et~al.}(2004)\citenamefont{Moon, Swift, and
  Swinney}}]{moon03}
\bibinfo{author}{\bibfnamefont{S.~J.} \bibnamefont{Moon}},
  \bibinfo{author}{\bibfnamefont{J.~B.} \bibnamefont{Swift}}, \bibnamefont{and}
  \bibinfo{author}{\bibfnamefont{H.~L.} \bibnamefont{Swinney}},
  \bibinfo{journal}{Phys. Rev. E} \textbf{\bibinfo{volume}{69}},
  \bibinfo{pages}{031301} (\bibinfo{year}{2004}).

\bibitem[{\citenamefont{Bizon et~al.}(1998)\citenamefont{Bizon, Shattuck,
  Swift, McCormick, and Swinney}}]{bizon98}
\bibinfo{author}{\bibfnamefont{C.}~\bibnamefont{Bizon}},
  \bibinfo{author}{\bibfnamefont{M.~D.} \bibnamefont{Shattuck}},
  \bibinfo{author}{\bibfnamefont{J.~B.} \bibnamefont{Swift}},
  \bibinfo{author}{\bibfnamefont{W.~D.} \bibnamefont{McCormick}},
  \bibnamefont{and} \bibinfo{author}{\bibfnamefont{H.~L.}
  \bibnamefont{Swinney}}, \bibinfo{journal}{Phys. Rev. Lett.}
  \textbf{\bibinfo{volume}{80}}, \bibinfo{pages}{57} (\bibinfo{year}{1998}).

\bibitem[{\citenamefont{Bougie et~al.}(2002)\citenamefont{Bougie, Moon, Swift,
  and Swinney}}]{bougie2002}
\bibinfo{author}{\bibfnamefont{J.}~\bibnamefont{Bougie}},
  \bibinfo{author}{\bibfnamefont{S.~J.} \bibnamefont{Moon}},
  \bibinfo{author}{\bibfnamefont{J.~B.} \bibnamefont{Swift}}, \bibnamefont{and}
  \bibinfo{author}{\bibfnamefont{H.~L.} \bibnamefont{Swinney}},
  \bibinfo{journal}{Phys. Rev. E} \textbf{\bibinfo{volume}{66}},
  \bibinfo{pages}{051301} (\bibinfo{year}{2002}).

\bibitem[{\citenamefont{Landau and Lifshitz}(1959)}]{landauandlifshitz1959}
\bibinfo{author}{\bibfnamefont{L.~D.} \bibnamefont{Landau}} \bibnamefont{and}
  \bibinfo{author}{\bibfnamefont{E.~M.} \bibnamefont{Lifshitz}},
  \emph{\bibinfo{title}{Fluid Mechanics}} (\bibinfo{publisher}{Pergamon Books
  Ltd., Oxford}, \bibinfo{year}{1959}).

\bibitem[{\citenamefont{Zaitsev and Shilomis}(1971)}]{zaitsev}
\bibinfo{author}{\bibfnamefont{V.~M.} \bibnamefont{Zaitsev}} \bibnamefont{and}
  \bibinfo{author}{\bibfnamefont{M.~I.} \bibnamefont{Shilomis}},
  \bibinfo{journal}{Soviet Physics JETP} \textbf{\bibinfo{volume}{32}},
  \bibinfo{pages}{866} (\bibinfo{year}{1971}).

\bibitem[{\citenamefont{Swift and Hohenberg}(1977)}]{swifthohenberg}
\bibinfo{author}{\bibfnamefont{J.~B.} \bibnamefont{Swift}} \bibnamefont{and}
  \bibinfo{author}{\bibfnamefont{P.~C.} \bibnamefont{Hohenberg}},
  \bibinfo{journal}{Phys. Rev. A} \textbf{\bibinfo{volume}{15}},
  \bibinfo{pages}{319} (\bibinfo{year}{1977}).

\bibitem[{\citenamefont{Wu et~al.}(1995)\citenamefont{Wu, Ahlers, and
  Cannell}}]{wu}
\bibinfo{author}{\bibfnamefont{M.}~\bibnamefont{Wu}},
  \bibinfo{author}{\bibfnamefont{G.}~\bibnamefont{Ahlers}}, \bibnamefont{and}
  \bibinfo{author}{\bibfnamefont{D.~S.} \bibnamefont{Cannell}},
  \bibinfo{journal}{Phys. Rev. Lett.} \textbf{\bibinfo{volume}{75}},
  \bibinfo{pages}{1743} (\bibinfo{year}{1995}).

\bibitem[{\citenamefont{Oh and Ahlers}(2003)}]{oh}
\bibinfo{author}{\bibfnamefont{J.}~\bibnamefont{Oh}} \bibnamefont{and}
  \bibinfo{author}{\bibfnamefont{G.}~\bibnamefont{Ahlers}},
  \bibinfo{journal}{Phys. Rev. Lett.} \textbf{\bibinfo{volume}{91}},
  \bibinfo{pages}{094501} (\bibinfo{year}{2003}).

\bibitem[{\citenamefont{Goldman et~al.}(2004)\citenamefont{Goldman, Swift, and
  Swinney}}]{goldman2004}
\bibinfo{author}{\bibfnamefont{D.~I.} \bibnamefont{Goldman}},
  \bibinfo{author}{\bibfnamefont{J.~B.} \bibnamefont{Swift}}, \bibnamefont{and}
  \bibinfo{author}{\bibfnamefont{H.~L.} \bibnamefont{Swinney}},
  \bibinfo{journal}{Phys. Rev. Lett.} \textbf{\bibinfo{volume}{92}},
  \bibinfo{pages}{174302} (\bibinfo{year}{2004}).

\bibitem[{\citenamefont{Brey et~al.}(2009)\citenamefont{Brey, Maynar, and
  de~Soria}}]{brey2009}
\bibinfo{author}{\bibfnamefont{J.~J.} \bibnamefont{Brey}},
  \bibinfo{author}{\bibfnamefont{P.}~\bibnamefont{Maynar}}, \bibnamefont{and}
  \bibinfo{author}{\bibfnamefont{M.~I.~G.} \bibnamefont{Garc\'{i}a~de~Soria}},
  \bibinfo{journal}{Phys. Rev. E.} \textbf{\bibinfo{volume}{79}},
  \bibinfo{pages}{051305} (\bibinfo{year}{2009}).

\bibitem[{\citenamefont{Bougie}(2010)}]{epaps}
  \bibinfo{journal}{See supplementary material at 
http://link.aps.org/supplemental/10.1103/PhysRevE.81.032301 
for a list of equations used.} 

\bibitem[{\citenamefont{Melo and Douady}(1993)}]{melo1993}
\bibinfo{author}{\bibfnamefont{F.}~\bibnamefont{Melo}} \bibnamefont{and}
  \bibinfo{author}{\bibfnamefont{S.}~\bibnamefont{Douady}},
  \bibinfo{journal}{Phys. Rev. Lett.} \textbf{\bibinfo{volume}{71}},
  \bibinfo{pages}{3283} (\bibinfo{year}{1993}).

\bibitem[{\citenamefont{Umbanhowar and Swinney}(2000)}]{umbanhowar2000}
\bibinfo{author}{\bibfnamefont{P.}~\bibnamefont{Umbanhowar}} \bibnamefont{and}
  \bibinfo{author}{\bibfnamefont{H.~L.} \bibnamefont{Swinney}},
  \bibinfo{journal}{Physica A} \textbf{\bibinfo{volume}{288}},
  \bibinfo{pages}{344} (\bibinfo{year}{2000}).

\end{thebibliography}

\end{document}